\documentclass[]{spie}  %>>> use for US letter paper
\usepackage{amsmath,amsfonts,amssymb}
\usepackage{graphicx}
\usepackage[colorlinks=true, allcolors=blue]{hyperref}
\usepackage{aas_macros}

\title{CATKit2-HCI: a collaborative framework for advancing high-contrast coronagraph testbeds}

\author[a]{R\'emi Soummer}
\author[b,c,d]{Iva Laginja}
\author[ ]{Emiel H. Por}
\author[e]{Rapha\"el Pourcelot}
\author[a]{Sarah Steiger}
\author[a]{Bryony Nickson}
\author[f]{Alexis Lau}
\author[c]{Pierre Baudoz}
\author[g]{Clarissa Do \'O}
\author[h]{Jules Fowler}
\author[c]{Rapha\"el Galicher}
\author[i]{Tyler D. Groff}
\author[h]{Rebecca Jensen-Clem}
\author[g]{Dimitri P. Mawet}
\author[c]{Johan Mazoyer}
\author[i]{Michael W. McElwain}
\author[i]{Lane Meier}
\author[c]{Axel Potier}
\author[a]{Laurent A. Pueyo}
\author[j]{Susan F. Redmond}
\author[i]{Scott D. Will}
\author[i]{Neil T. Zimmerman}

\affil[a]{Space Telescope Science Institute, 3700 San Martin Drive, Baltimore, MD 21218, USA}
\affil[b]{Universit\'e C\^ote d'Azur, Observatoire de la C\^ote d'Azur, CNRS, Laboratoire Lagrange, Nice, France}
\affil[c]{LIRA, Observatoire de Paris, Université PSL, Sorbonne Université, Université Paris Cité, CY Cergy Paris Université, CNRS, 92195 Meudon, France}
\affil[d]{NOVA/Leiden University, Einsteinweg 55, 2333 CC Leiden, The Netherlands}
\affil[e]{Max Planck Institute for Astronomy, Heidelberg, Germany}
\affil[f]{Aix Marseille Université, CNRS, CNES, LAM, Marseille, France}
\affil[g]{California Institute of Technology, Pasadena, CA, USA}
\affil[h]{Department of Astronomy and Astrophysics, University of California, Santa Cruz, CA, USA}
\affil[i]{NASA Goddard Space Flight Center, Greenbelt, MD, USA}
\affil[j]{Jet Propulsion Laboratory, California Institute of Technology, Pasadena, CA, USA}

\authorinfo{Further author information, send correspondence to R.S.: E-mail: soummer@stsci.edu}

\begin{document}
\maketitle

\begin{abstract}
High-contrast exoplanet imaging requires dedicated laboratory testbeds for the development and validation of coronagraph architectures, wavefront sensing and control methods, calibration strategies, and system-level observing concepts. These testbeds often share similar software needs, yet many tools are developed independently at each institution. The CATKit2-High-Contrast-Imaging collaboration, or CATKit2-HCI, addresses this gap by providing a shared software framework for reusable HCI infrastructure. Built on top of CATKit2, an open-source hardware control and synchronization framework originally developed for the High-contrast Imager for Complex Aperture Telescopes (HiCAT) testbed at the Space Telescope Science Institute, CATKit2-HCI provides the collaborative layer for HCI-specific algorithms, calibration tools, diagnostics, visualization, and performance metrics. The collaboration currently includes multiple coronagraph testbeds in the United States and Europe. Its goals are to reduce duplicated software development, improve code quality through shared review, enable more direct comparison of results across facilities, and facilitate the movement of students, postdoctoral researchers, and collaborators between laboratories. We describe the motivation, architecture, collaboration model, shared technical capabilities, and early cross-testbed examples of CATKit2-HCI as a framework for accelerating coronagraph technology development.
\end{abstract}


\section{INTRODUCTION}
\label{sec:introduction}

\subsection{Scientific context and problem statement}
\label{sec:context}

Direct imaging and spectroscopy of exoplanets at high contrast requires the coordinated development of advanced coronagraphs, wavefront sensing and control algorithms, deformable mirror (DM) technologies, precision calibration procedures, and stable observing strategies. Laboratory testbeds play a central role in this development path by providing controlled environments in which coronagraph architectures can be assembled, calibrated, operated, and validated before similar concepts are considered for astronomical facilities or future space missions.

This laboratory infrastructure is particularly important for future exoplanet imaging missions such as the Habitable Worlds Observatory (HWO)\cite{2024SPIE13092E..1NF, 2026ASPC..543D..12P}, for which starlight suppression must be demonstrated under increasingly realistic conditions. Similar challenges also arise for ground-based coronagraphs on current and future large telescopes, including the European Extremely Large Telescope (ELT).

Although individual facilities differ in aperture geometry, coronagraph architecture, DM configuration, environmental stability, wavelength range, and science goals, much of the software architecture is not fundamentally testbed-specific. High-contrast imaging (HCI) testbed teams routinely acquire and process calibration data, define dark zones, estimate focal-plane electric fields, control DMs, monitor hardware states, compute contrast metrics, and generate diagnostic plots. As the number and sophistication of coronagraph testbeds increase, shared software infrastructure becomes an enabling technology for laboratory astrophysics: common tools, interfaces, and diagnostics can accelerate progress, while their absence forces each laboratory to rebuild similar capabilities independently.

\subsection{CATKit2 foundation}
\label{sec:catkit2}

The CATKit2 package\cite{Catkit2} provides the public hardware and service-control foundation on which CATKit2-HCI is built. It was originally developed in the context of the High-contrast Imager for Complex Aperture Telescopes (HiCAT\cite{Soummer2024HiCAT}) at the Space Telescope Science Institute (STScI). Its purpose is to provide a testbed-agnostic infrastructure for laboratory hardware control, including device communication, synchronization, shared memory, configuration management, and experiment orchestration.

The core design of CATKit2 is intentionally general. It provides the lower-level infrastructure needed to operate devices such as cameras, DMs, motion stages, light sources, and other laboratory hardware through a service-based architecture. This separation between hardware services and experiment logic allows testbed-specific packages to interact with devices through common service abstractions rather than ad hoc hardware scripts.

\subsection{The missing layer}
\label{sec:missing-layer}

The missing layer is the software space between hardware control and private, testbed-specific science experiments. This layer includes standard HCI algorithms, calibration workflows, optical modeling utilities, diagnostics, visualization tools, performance metrics, and data products that are commonly useful across multiple coronagraphic testbeds. These tools are not raw hardware drivers, but they are also not necessarily unpublished research experiments; they form the reusable laboratory infrastructure that supports daily HCI testbed operations.

Before CATKit2-HCI, this layer was typically developed locally within each testbed repository, which made mature tools difficult to share, review, and compare. CATKit2-HCI was created to provide a common collaborative layer above CATKit2 and below the private repositories of individual testbeds.

\section{CATKIT2-HCI COLLABORATION ARCHITECTURE AND MODEL}
\label{sec:architecture}

\subsection{Overall concept}
\label{sec:overall-concept}

CATKit2-HCI is a privately shared software framework for reusable HCI infrastructure. The central idea is to separate generic laboratory capabilities from local experiments, bench-specific tuning, and unpublished science demonstrations. Each testbed retains its own repository, operating procedures, and freedom to develop new research ideas, while mature tools useful beyond a single bench can be generalized, reviewed, and proposed for migration into CATKit2-HCI.

The goal is not to force every laboratory to operate identically. Coronagraph testbeds are diverse by design, and that diversity is scientifically valuable. CATKit2-HCI instead targets the parts of the software stack where needs naturally overlap, including standard algorithms, calibration and analysis utilities, file-output conventions, plotting routines, monitoring tools, and reusable abstractions for wavefront control experiments.

\subsection{Three-layer architecture}
\label{sec:three-layer}

\begin{figure}[th!]
\centering
\includegraphics[width=0.5\textwidth]{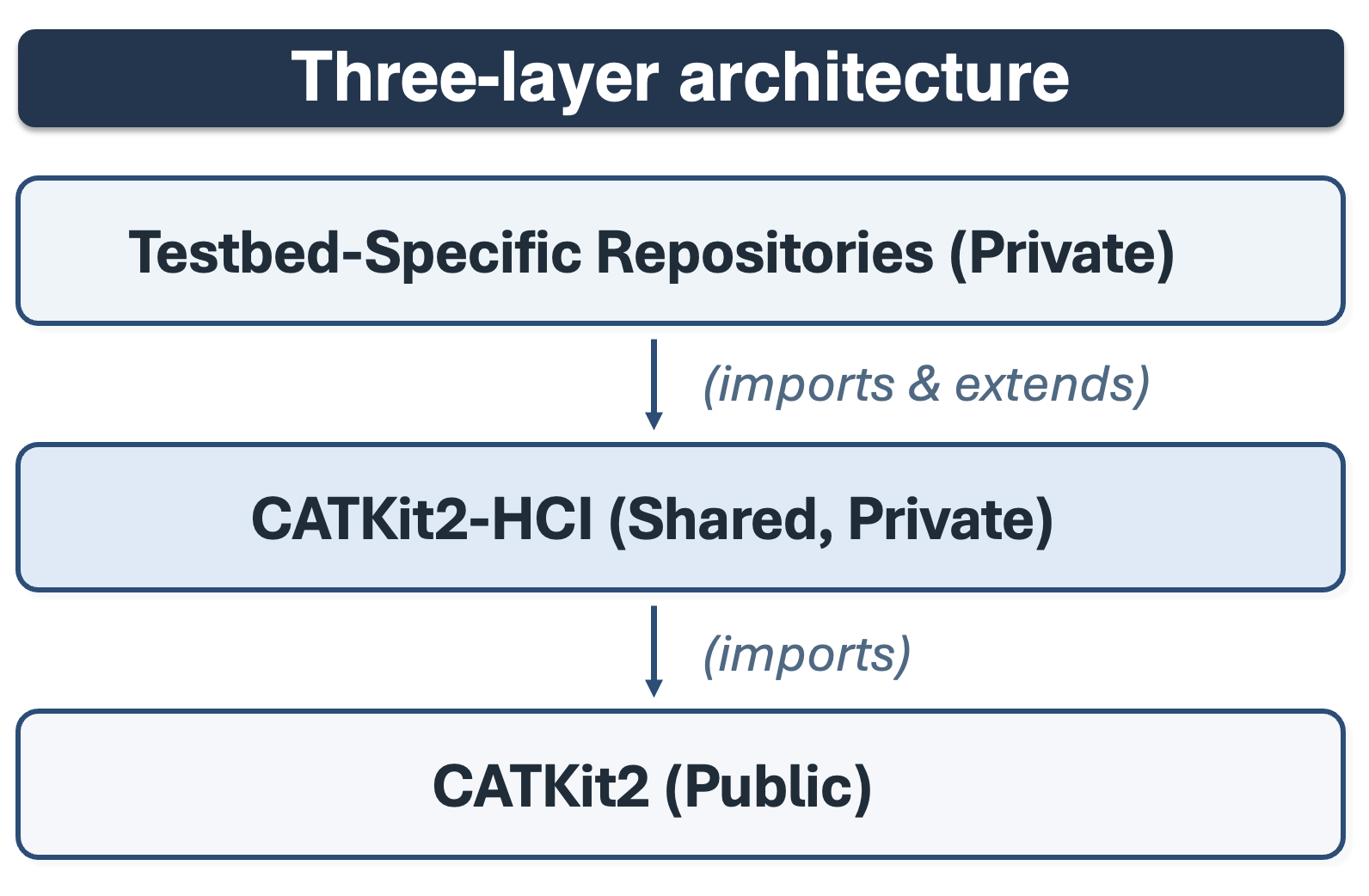}
\caption{\small Three-layer architecture of CATKit2, CATKit2-HCI, and private testbed repositories. See text for details.}
\label{fig:three_layer_architecture}
\end{figure}

The CATKit2-HCI collaboration is organized around a three-layer architecture, illustrated in Fig.~\ref{fig:three_layer_architecture}. The bottom layer is CATKit2, the public hardware and service-control infrastructure for device communication, synchronization, shared memory, configuration management, and experiment orchestration. The middle layer is CATKit2-HCI, the shared private collaboration repository containing reusable HCI infrastructure such as wavefront sensing and control algorithms, calibration and data tools, optical modeling utilities, diagnostics, plotting, visualization, and monitoring capabilities. The top layer consists of the private repositories of individual testbeds, which contain local experiments, bench-specific hardware adjustments, procedures, calibration products, unpublished science demonstrations, and configuration files.

This architecture defines a practical software boundary. CATKit2-HCI captures functionality common across HCI testbeds but too specialized to belong in the hardware-control layer. Private repositories import and extend CATKit2-HCI when appropriate while preserving the autonomy needed for each laboratory to pursue its own research program.

\subsection{Collaboration model}
\label{sec:collaboration-model}

The CATKit2-HCI collaboration is designed for active participation rather than passive access. The collaboration agreement asks each testbed team to contribute reusable tools, participate in technical discussions and reviews, and migrate mature, generic functionality from local repositories into the shared CATKit2-HCI layer when appropriate. It also establishes a reciprocal ``rebasing'' norm: when a capability has been generalized and merged into CATKit2-HCI, participating testbeds are expected to refactor equivalent local code to use the shared implementation. This principle is not intended primarily as an enforcement mechanism, but as a shared commitment to keep CATKit2-HCI technically coherent as a living common codebase rather than a set of parallel local forks.

\subsection{Participating testbeds}
\label{sec:participating-testbeds}

The collaboration currently includes six participating coronagraph testbeds listed in Tab.~\ref{tab:testbeds}: HiCAT\cite{Soummer2024HiCAT} and CAPSULE at STScI in Baltimore; THD2\cite{Baudoz2024PolarizationEffects} at LIRA / Observatoire de Paris in Meudon; SEAL\cite{10.1117/12.3065350} at the University of California, Santa Cruz; HCST\cite{2024SPIE13092E..6BB} at Caltech in Pasadena; and ExoSPEC\cite{2024SPIE13092E..5KM} at NASA Goddard Space Flight Center in Greenbelt.

These testbeds span different institutions, optical architectures, and research goals. Their participation provides an opportunity to test whether shared HCI software abstractions can support a range of laboratory environments, while also providing a practical mechanism for transferring software knowledge between teams.

\begin{table}[th!]
\caption{Participating testbeds in the CATKit2-HCI collaboration as of July 2026.}
\label{tab:testbeds}
\centering
\begin{tabular}{lll}
\hline
Testbed & Institution & Location \\
\hline
HiCAT & STScI & Baltimore, MD, USA \\
CAPSULE & STScI & Baltimore, MD, USA \\
THD2 & LIRA / Observatoire de Paris & Meudon, France \\
SEAL & UC Santa Cruz & Santa Cruz, CA, USA \\
HCST & Caltech & Pasadena, CA, USA \\
ExoSPEC & NASA GSFC & Greenbelt, MD, USA \\
\hline
\end{tabular}
\end{table}

\section{SHARED TECHNICAL CAPABILITIES AND EARLY EXAMPLES}
\label{sec:technical-capabilities}

The framework is under active development, with mature routines being generalized from existing testbed repositories and new tools being developed directly in the shared layer. A major focus is a common software architecture for wavefront sensing and control. Many high-contrast testbeds rely on related concepts, including focal-plane wavefront estimation, probe-based electric-field sensing, and DM control. CATKit2-HCI provides a place to implement standard or classical algorithms in a reusable form, including electric field conjugation\cite{2007EFC} (EFC), pairwise probing\cite{2011SPIE.8151E..10G}, differential Optical Transfer Function phase retrieval \cite{Codona2013dOTF,2023SPIE12680E..2HN}, dark-zone definitions, and performance observers.

The control framework separates estimators, controllers, observers, and testbed-specific interfaces. Estimators convert image data and calibration information into quantities such as focal-plane electric-field or phase estimates. Controllers convert those estimates into commands for DMs or other actuators. Observers generate diagnostics, performance metrics, plots, and standardized outputs during or after an experiment. This abstraction makes it easier to replace one component without rewriting a full experiment, and helps different testbeds compare algorithmic choices because the same estimator or observer can be exercised in different optical environments.

CATKit2-HCI also addresses recurring needs in DM operations, data calibration, modeling, and visualization. DM utilities include actuator maps, command formatting, multi-DM abstractions, command-history storage, and readable output products. Data and calibration tools include dark calibration, auto-exposure, spot finding, image registration, contrast metrics, and image-based sampling estimates. Optical-modeling utilities include HCIPy\cite{Por2018HighContrastImaging}-based models, propagation helpers, standardized plotting routines, and tools for comparing modeled and measured behavior. The graphical user interface (GUI) and dashboard infrastructure supports real-time inspection of services, images, scalar time series, logs, experiments, and configurable dashboard panels, while allowing each laboratory to retain testbed-specific panels and configurations.

\begin{figure}[th!]
\centering
\includegraphics[width=1.0\textwidth]{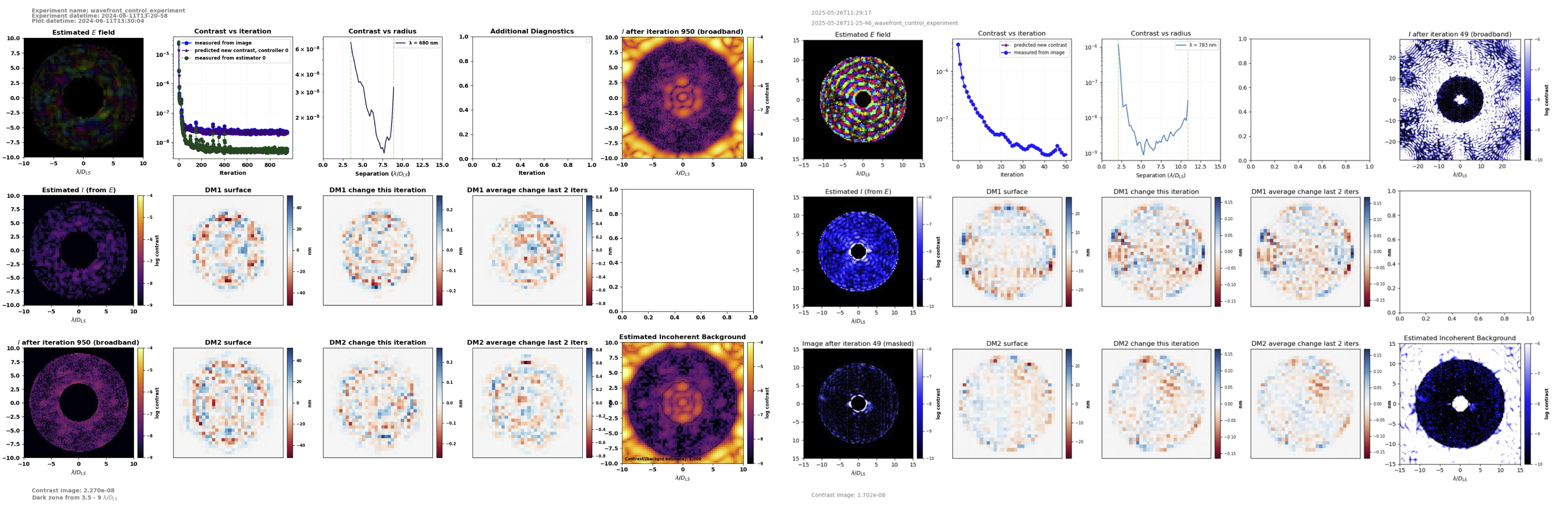}
\caption{\small Shared EFC plotting and analysis outputs for HiCAT (left) and THD2 (right).}
\label{fig:efc_comparison}
\end{figure}

Two early examples illustrate the intended use of this shared layer. The first is the use of common plotting and analysis routines for EFC\cite{2007EFC} experiments on independent testbeds, shown in Fig.~\ref{fig:efc_comparison}. HiCAT and THD2 have different optical architectures and local configurations, but they share many of the same needs when evaluating wavefront control experiments: tracking contrast evolution, visualizing dark-zone performance, monitoring convergence behavior, recording DM commands, and producing diagnostic plots.

\begin{figure}[th!]
\centering
\includegraphics[width=1.0\textwidth]{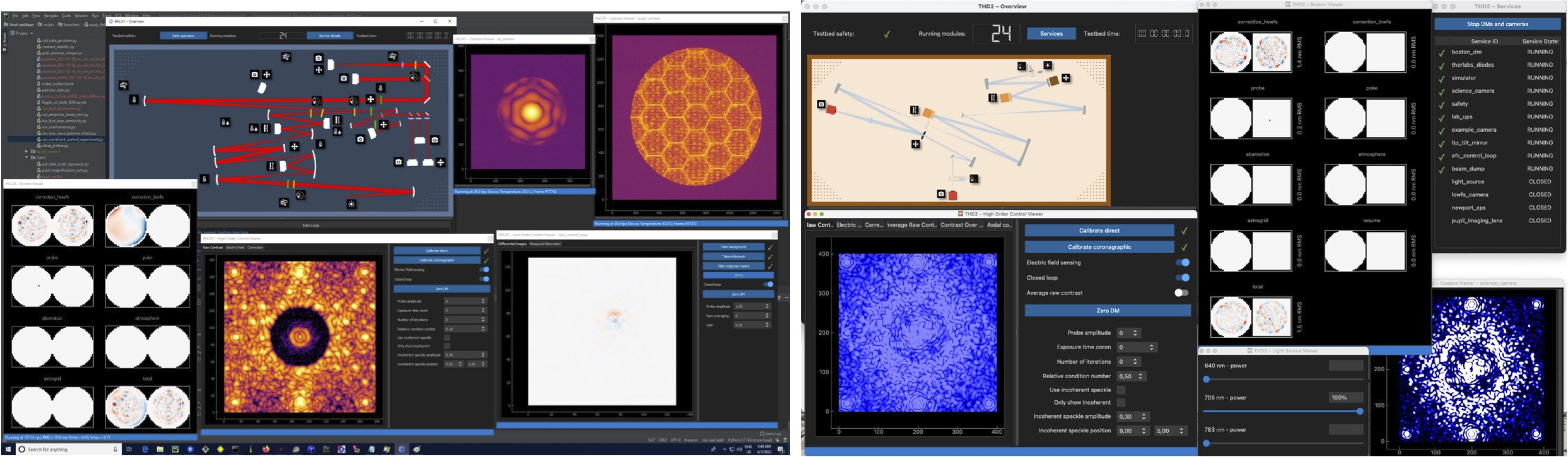}
\caption{\small CATKit2-HCI GUI/dashboard concepts applied to HiCAT (left) and THD2 (right).}
\label{fig:gui_dashboard}
\end{figure}

The second example is the development of shared GUI and dashboard concepts, shown in Fig.~\ref{fig:gui_dashboard}. Laboratory testbeds require constant monitoring of hardware services, images, scalar values, logs, and experiment state. Historically, these tools are often built locally and can become tightly coupled to a single bench. CATKit2-HCI provides a path toward reusable monitoring components that can be configured for different testbeds. The GUI framework also reinforces a shared operational vocabulary: when teams use similar concepts for services, properties, image streams, scalar histories, and experiment execution, discussions become more efficient. The benefit is shared operational understanding while preserving local testbed configurations.

Together, these capabilities reduce duplicated code, improve review, support common diagnostics, facilitate algorithm transfer, and help new contributors move more easily between laboratories. In this sense, CATKit2-HCI is both a software project and a community infrastructure project.

\section{CONCLUSIONS}
\label{sec:conclusions}

CATKit2-HCI fills the software layer between public CATKit2 hardware infrastructure and private testbed-specific experiment repositories. By collecting mature, reusable HCI tools in a shared collaboration repository, it reduces duplicated effort while preserving the flexibility needed for independent laboratory research. Its three-layer architecture and rebasing principle are intended to keep the shared codebase technically coherent without forcing all testbeds to operate identically.

Early examples from HiCAT and THD2 show how common EFC diagnostics and GUI/dashboard concepts can support cross-testbed comparison, training, and collaboration. As the framework matures, CATKit2-HCI aims to establish sustainable, community-vetted software patterns for HCI testbeds and to accelerate coronagraph technology development for future exoplanet imaging facilities, including HWO and advanced ground-based systems.

\acknowledgments

This work benefited from the CATKit2 high-contrast imaging collaboration which originated at the HiCAT testbed at STScI. HiCAT was developed over the past 10 years and benefitted from the work of an extended collaboration of over 50 people and supported in part by the National Aeronautics and Space Administration under Grant 80NSSC19K0120 issued through the Strategic Astrophysics Technology/Technology Demonstration for Exo-planet Missions Program (SAT-TDEM; PI: R. Soummer), and under Grant 80NSSC22K0372 issued through the Astrophysics Research and Analysis Program (APRA; PI: L. Pueyo).
The development of the THD2 test bench was partly supported by Centre National d'\'Etudes Spatiales (CNES) through R\&D fundings R-S14/SU-002-068, R-S17/SU-0002-068, R-S19/SU-0002-105.
Emiel H. Por was supported in part by the NASA Hubble Fellowship grant HST-HF2-51467.001-A awarded by the Space Telescope Science Institute, which is operated by the Association of Universities for Research in Astronomy, Incorporated, under NASA contract NAS5-26555.
Sarah Steiger acknowledges support from an STScI Postdoctoral Fellowship.  
Iva Laginja acknowledges partial support from a postdoctoral fellowship issued by the Centre National d’Etudes Spatiales (CNES) in France.
Iva Laginja acknowledges partial support by the European Space Agency (ESA) under the tender number TDE-TEC-MOO AO/1-11613/23/NL/AR in the context of the SUPPPPRESS project.
The authors acknowledge the use of ChatGPT for language and grammar improvements and cleanup. 

% References
\bibliography{references}

@INPROCEEDINGS{2011SPIE.8151E..10G,
       author = {{Give'on}, Amir and {Kern}, Brian D. and {Shaklan}, Stuart},
        title = "{Pair-wise, deformable mirror, image plane-based diversity electric field estimation for high contrast coronagraphy}",
    booktitle = {Techniques and Instrumentation for Detection of Exoplanets V},
         year = 2011,
       editor = {{Shaklan}, Stuart},
       series = {Society of Photo-Optical Instrumentation Engineers (SPIE) Conference Series},
       volume = {8151},
        month = oct,
          eid = {815110},
        pages = {815110},
          doi = {10.1117/12.895117},
       adsurl = {https://ui.adsabs.harvard.edu/abs/2011SPIE.8151E..10G}
}

@INPROCEEDINGS{2024SPIE13092E..1NF,
       author = {{Feinberg}, Lee and {Ziemer}, John and {Ansdell}, Megan and {Crooke}, Julie and {Dressing}, Courtney and {Mennesson}, Bertrand and {O'Meara}, John and {Pepper}, Joshua and {Roberge}, Aki},
        title = "{The Habitable Worlds Observatory engineering view: status, plans, and opportunities}",
    booktitle = {Space Telescopes and Instrumentation 2024: Optical, Infrared, and Millimeter Wave},
         year = 2024,
       editor = {{Coyle}, Laura E. and {Matsuura}, Shuji and {Perrin}, Marshall D.},
       series = {Society of Photo-Optical Instrumentation Engineers (SPIE) Conference Series},
       volume = {13092},
        month = aug,
          eid = {130921N},
        pages = {130921N},
          doi = {10.1117/12.3018328},
       adsurl = {https://ui.adsabs.harvard.edu/abs/2024SPIE13092E..1NF}
}

@INPROCEEDINGS{2026ASPC..543D..12P,
       author = {{Postman}, Marc and {Stapelfeldt}, Karl},
        title = "{Foreword: The Habitable Worlds Observatory in Historical Context}",
    booktitle = {Astronomical Society of the Pacific Conference Series},
         year = 2026,
       editor = {{Lee}, Janice C. and {Noviello}, Jessica and {LaMassa}, Stephanie and {Postman}, Marc},
       series = {Astronomical Society of the Pacific Conference Series},
       volume = {543},
        month = nov,
        pages = {xii},
       adsurl = {https://ui.adsabs.harvard.edu/abs/2026ASPC..543D..12P}
}

@INPROCEEDINGS{2024SPIE13092E..5KM, author = {{McElwain}, Michael W. and {Zimmerman}, Neil T. and {Rauscher}, Bernard and {Groff}, Tyler and {Mandell}, Avi and {Alei}, Eleonora and {Baines}, Tyler and {Berrier}, Joshua and {Bradley}, Harrison and {Gong}, Qian and {Howe}, Alex and {Juanola-Parramon}, Roser and {Kan}, Emily and {Kelly}, Daniel and {Kofman}, Vincent and {Sitarski}, Breann and {Stark}, Christopher and {Subedi}, Hari and {Villanueva}, Geronimo and {Will}, Scott}, title = "{ExoSpec project: exoplanet spectroscopy technologies for the Habitable Worlds Observatory at NASA's Goddard Space Flight Center}", booktitle = {Space Telescopes and Instrumentation 2024: Optical, Infrared, and Millimeter Wave}, year = 2024, editor = {{Coyle}, Laura E. and {Matsuura}, Shuji and {Perrin}, Marshall D.}, series = {Society of Photo-Optical Instrumentation Engineers (SPIE) Conference Series}, volume = {13092}, month = aug, eid = {130925K}, pages = {130925K}, doi = {10.1117/12.3020211}, adsurl = {https://ui.adsabs.harvard.edu/abs/2024SPIE13092E..5KM} }

@inproceedings{Baudoz2024PolarizationEffects,
author = {Pierre Baudoz and Celia Desgrange and Rapha{\"e}l Galicher and Iva Laginja},
title = {{Polarization effects on high contrast imaging: measurements on THD2 bench}},
volume = {13092},
booktitle = {Space Telescopes and Instrumentation 2024: Optical, Infrared, and Millimeter Wave},
editor = {Laura E. Coyle and Shuji Matsuura and Marshall D. Perrin},
publisher = {SPIE},
pages = {130926L},
year = {2024},
doi = {10.1117/12.3020010},
URL = {https://doi.org/10.1117/12.3020010}
}

@inproceedings{Soummer2024HiCAT,
author = {R{\'e}mi Soummer and Rapha{\"e}l Pourcelot and Emiel H. Por and Sarah Steiger and Iva Laginja and Benjamin Buralli and Susan Redmond and Laurent Pueyo and Marshall D. Perrin and Marc Ferrari and Jules Fowler and John Hagopian and Mamadou N'Diaye and Meiji Nguyen and Bryony Nickson and Peter Petrone and Ananya Sahoo and Anand Sivaramakrishnan and Scott D. Will},
title = {{High-contrast imager for complex aperture telescopes (HiCAT): 11. System-level demonstration of the apodized pupil Lyot coronagraph with a segmented aperture in air}},
volume = {13092},
booktitle = {Space Telescopes and Instrumentation 2024: Optical, Infrared, and Millimeter Wave},
editor = {Laura E. Coyle and Shuji Matsuura and Marshall D. Perrin},
publisher = {SPIE},
pages = {130921Z},
year = {2024},
doi = {10.1117/12.3018037},
URL = {https://doi.org/10.1117/12.3018037}
}

@inproceedings{10.1117/12.3065350,
author = {Rebecca Jensen-Clem and Vincent Chambouleyron and Prince Javier and Daren Dillon and Emiel H. Por and Benjamin Calvin and Sylvain Cetre and Rodrigo Amezcua Correa and Tara Crowe and Jordan Diaz and Caleb Dobias and David Doelman and Stephen Eikenberry and J. Fowler and Benjamin L. Gerard and Phil Hinz and Renate Kupke and Ashai Moreno and Tiffany Nguyen and Maissa Salama and Aditya R. Sengupta and Nour Skaf and Frans Snik},
title = {{The Santa Cruz Extreme AO Lab (SEAL) 2.0: a reflective, multiwavelength rebuild}},
volume = {13627},
booktitle = {Techniques and Instrumentation for Detection of Exoplanets XII},
editor = {Garreth J. Ruane and Maxwell A. Millar-Blanchaer},
organization = {International Society for Optics and Photonics},
publisher = {SPIE},
pages = {1362727},
year = {2025},
doi = {10.1117/12.3065350},
URL = {https://doi.org/10.1117/12.3065350}
}

@INPROCEEDINGS{2024SPIE13092E..6BB,
       author = {{Bertrou-Cantou}, A. and {Redmond}, S. and {Mawet}, D. and {Sercel}, G. and {Echeverri}, D. and {Desai}, N. and {Llop-Sayson}, J. and {Ruane}, G. and {Serabyn}, E. and {Wallace}, J.~K.},
        title = "{High-contrast spectroscopy testbed (HCST): tip/tilt sensing in reflection of the vector vortex coronagraph (VVC)}",
    booktitle = {Space Telescopes and Instrumentation 2024: Optical, Infrared, and Millimeter Wave},
         year = 2024,
       editor = {{Coyle}, Laura E. and {Matsuura}, Shuji and {Perrin}, Marshall D.},
       series = {Society of Photo-Optical Instrumentation Engineers (SPIE) Conference Series},
       volume = {13092},
        month = aug,
          eid = {130926B},
        pages = {130926B},
          doi = {10.1117/12.3019237},
       adsurl = {https://ui.adsabs.harvard.edu/abs/2024SPIE13092E..6BB}
}

@INPROCEEDINGS{2007EFC,
       author = {{Give'on}, Amir and {Kern}, Brian and {Shaklan}, Stuart and {Moody}, Dwight C. and {Pueyo}, Laurent},
        title = "{Broadband wavefront correction algorithm for high-contrast imaging systems}",
    booktitle = {Astronomical Adaptive Optics Systems and Applications III},
         year = 2007,
       editor = {{Tyson}, Robert K. and {Lloyd-Hart}, Michael},
       series = {Society of Photo-Optical Instrumentation Engineers (SPIE) Conference Series},
       volume = {6691},
        month = sep,
          eid = {66910A},
        pages = {66910A},
          doi = {10.1117/12.733122},
       adsurl = {https://ui.adsabs.harvard.edu/abs/2007SPIE.6691E..0AG}
}

@misc{Catkit2,
  author       = {Emiel H. Por and
                  Iva Laginja and
                  Sarah Steiger and
                  Rémi Soummer and
                  Lane Meier and
                  Erin Pougheon and
                  Raphaël Pourcelot and
                  Michael Paul Philbin and
                  Johan Mazoyer and
                  Arnaud Sevin and
                  Meiji Nguyen and
                  Ananya Sahoo and
                  Alexis Lau and
                  Corentin Paviot and
                  Léo Egger and
                  Lukas Delaye and
                  Alex Meredith and
                  P. L. Lim and
                  linarphy and
                  Jules Fowler},
  title        = {spacetelescope/catkit2: v0.7.0. [\textit{Zenodo}] 10.5281/zenodo.20843325},
  month        = jun,
  year         = 2026,
  publisher    = {Zenodo},
  version      = {v0.7.0},
  doi          = {10.5281/zenodo.20843325},
  url          = {https://doi.org/10.5281/zenodo.20843325},
}

@INPROCEEDINGS{2023SPIE12680E..2HN,
       author = {{Nguyen}, Meiji M. and {Por}, Emiel H. and {Sahoo}, Ananya and {Soummer}, R{\'e}mi and {Pourcelot}, Rapha{\"e}l. and {Nickson}, Bryony F. and {Pueyo}, Laurent and {Perrin}, Marshall and {Laginja}, Iva},
        title = "{Differential optical transfer function wavefront sensing and pupil calibration with the HiCAT testbed}",
    booktitle = {Society of Photo-Optical Instrumentation Engineers (SPIE) Conference Series},
         year = 2023,
       series = {Society of Photo-Optical Instrumentation Engineers (SPIE) Conference Series},
       volume = {12680},
        month = oct,
          eid = {126802H},
        pages = {126802H},
          doi = {10.1117/12.2677756},
       adsurl = {https://ui.adsabs.harvard.edu/abs/2023SPIE12680E..2HN}
}

@inproceedings{Por2018HighContrastImaging,  
author = {Emiel H. Por and Sebastiaan Y. Haffert and Vikram M. Radhakrishnan and David S. Doelman and Maaike van Kooten and Steven P. Bos},  
title = {{High Contrast Imaging for Python (HCIPy): an open-source adaptive optics and coronagraph simulator}},  
volume = {10703},  
booktitle = {Adaptive Optics Systems VI},  
editor = {Laird M. Close and Laura Schreiber and Dirk Schmidt}, 
publisher = {SPIE},  
pages = {1112 -- 1125},  
year = {2018},  
doi = {10.1117/12.2314407},  
URL = {https://doi.org/10.1117/12.2314407}  
}

@ARTICLE{Codona2013dOTF,
       author = {{Codona}, Johanan L.},
        title = "{Differential optical transfer function wavefront sensing}",
      journal = {Optical Engineering},
         year = 2013,
        month = sep,
       volume = {52},
          eid = {097105},
        pages = {097105},
          doi = {10.1117/1.OE.52.9.097105},
       adsurl = {https://ui.adsabs.harvard.edu/abs/2013OptEn..52i7105C}
}
\bibliographystyle{spiebib}

\end{document}